\documentclass[prd,aps]{revtex4}
\usepackage{amssymb,amsmath,amsfonts,amsbsy,epsfig}
\usepackage{slashed}

\newcommand{\be}{\begin{equation}}
\newcommand{\ee}{\end{equation}}
\newcommand{\bea}{\begin{eqnarray}}
\newcommand{\eea}{\end{eqnarray}}
\newcommand{\nn}{\nonumber}

\def\L{{\mathcal L}}
\def\M{{\mathcal M}}

\begin{document}

\title{Photon-Photon Scattering in Very Special Relativity}
\date{\today}

\author{Jorge Alfaro}
\email[]{jalfaro@fis.uc.cl}
\author{Alex Soto}
\email[]{arsoto1@uc.cl}
\affiliation{
Instituto de F\'{i}sica, Pontificia Universidad de Cat\'olica de Chile, \mbox{Av. Vicu\~na Mackenna 4860, Santiago, Chile}
}

\begin{abstract} 

Starting from the Mandelstam-Leibbrandt prescription, we introduce a general rule for the null vector $\bar{n}$ to compute any $SIM(2)$ integral and diagram with an arbitrary number of external legs. Using the new prescription, the computation of the low energy limit of photon-photon scattering under the Very Special Relativity (VSR) framework is presented. The prescription preserves automatically the Ward identities corresponding to the gauge symmetry. Within the low momentum approximation we get the standard unpolarized differential cross section for photon-photon scattering. The result suggests that loops with any external photon legs on-shell will be zero in VSR.


\end{abstract}

\maketitle 

\section{Introduction}

Photon-Photon scattering is a paradigmatic phenomenon predicted in quantum electrodynamics (QED). The first insight was given by Halpern, who proposed qualitatively that virtual electron-positron pairs could generate photon-photon collisions\cite{Halpern:1933dya}. The low energy calculation was developed by Euler and Kockel\cite{Euler:1935zz, Euler:1936oxn}, while the high energy scattering was considered in \cite{Akhiezer:1936vzu}. The low energy computation can be recovered from the Euler-Heisenberg lagrangian\cite{Heisenberg:1935qt} expanding to the quartic order in the constant field. Another way to proceed is to follow the Schwinger method using the so called proper-time formalism\cite{Schwinger:1951nm}. Furthermore, new analysis using Feynman diagrams were carried later in \cite{Karplus:1950zza,Karplus:1950zz}. Early results can be summarized in \cite{Costantini:1971cj} and an interesting review about historical point of view and some experiments developed could be found in \cite{Scharnhorst:2017wzh}.\\

 Standard QED rests on the Lorentz invariance. However, the observation of Cohen and Glashow that the universe could be invariant under a Lorentz subgroup instead of the full group\cite{Cohen:2006ky}, is an interesting approach that can explain the neutrino mass without neither new particles nor leptonic number violation\cite{Cohen:2006ir}. In this proposal, known as Very Special Relativity (VSR), the usual symmetry group is $SIM(2)$, which does not have invariant tensors besides the usual Lorentz invariant tensors. Moreover, all the standard features of Special Relativity, as addition of velocities or time dilation, are present. The main feature in the model is the privileged direction given by a null vector $n=(1,0,0,1)$ which transform as $n\to e^\phi n$. It allows the existence of new terms in the Lagrangian that contains the same number of $n$ in the numerator and denominator. With this, the neutrino equation reads

\be
\label{neutrinoeq}
\left(i\slashed{\partial}+\frac{1}{2}m^2\frac{\slashed{n}}{n\cdot \partial}\right)\nu=0,
\ee
where $m$ is the neutrino mass. From (\ref{neutrinoeq}) we get the dispersion relation $p^2=m^2$. Thus, we get the neutrino mass with the price of a non-local term introduced.\\

This model has been studied in the electroweak sector\cite{Alfaro:2015fha} and electrodynamics\cite{Cheon:2009zx}. The phenomenology of the VSR-QED has been reviewed in the kaon and pion decay\cite{Nayak:2016zed}, Coulomb scattering and Bremsstrahlung\cite{Alfaro:2019koq} and the Compton and Bhabha scattering\cite{Bufalo:2019kea}. However, the photon-photon scattering has not been studied yet in the framework. The main difficulty is a right prescription in the computation of the integrals containing the non local term $(n\cdot p)^{-1}$. In the two leg computations, as in the vacuum polarization and electron self energy, has been used the Mandelstam-Leibbrandt prescription\cite{Mandelstam:1982cb,Leibbrandt:1983pj}, where a new null vector $\bar{n}$ is introduced. Nevertheless, this new vector breaks the $SIM(2)$ invariance. To restore the symmetry in \cite{Alfaro:2017umk} the method used was to write $\bar{n}$ as a linear combination of $n$ and the external momentum involved. This methodology should change in the photon-photon scattering computation, since the diagrams has four external legs. Thus, the way to write $\bar{n}$ in terms of the momenta involved is not clear. In this article we present a way to deal with diagrams with any legs and we apply it in the cross section computation of the photon-photon scattering.\\

The outline of the paper is as follows. In the section II we will review the VSR-QED sector and the Feynman rules. In section III we will use the Feynman rules to compute the photon-photon scattering. In section IV we will present a prescription to compute the integrals with non-local terms and we will use it to get the cross section. In section V we will compute the scattering using the Schwinger proper-time method to compare with the diagrammatic computation. Finally, in the section VI we will summarize and discuss our results.

\section{QED VSR Model}

Let $\mathcal{L}$ be the VSR Lagrangian for a electron and an electromagnetic field, given by
\be
\mathcal{L}=\bar{\psi}\left(i\left(\slashed{D}+\frac{1}{2}m^2\frac{\slashed{n}}{n\cdot D}\right)-M\right)\psi-\frac{1}{4}F_{\mu\nu}F^{\mu\nu},
\ee
where $n$ is the null vector given by the VSR theory, $m$ a mass parameter, which we associate with the neutrino mass, $\psi$ is the electron field, $F_{\mu\nu}=\partial_\mu A_\nu-\partial_\nu A_\mu$ and $D_\mu=\partial_\mu+ieA_\mu$. Although a photon mass is allowed in VSR, without loss of gauge invariance\cite{Alfaro:2013uva,Alfaro:2015fha,Alfaro:2019koq}, in the following steps we will neglect it, because its contribution is very small in comparison with $m^2$. Starting from the equation of motion for electron, the dispersion relation for the electron shows us the electron mass is given by $M_e=\sqrt{M^2+m^2}$.\\

Due to the non local nature of $1/(n\cdot D)$, we can get different kind of vertex between two fermion lines and photon lines. We keep only vertex up to four photonic legs, which are relevant in the photon-photon scattering computation. Thus, we expand up to $e^4$ and we get

\bea
\label{lagrangianqedvsr}
\mathcal{L}&=&\bar{\psi}\left(i\slashed{\partial}-e\slashed{A}+\frac{i}{2}m^2\frac{\slashed{n}}{n\cdot\partial}\left(1-i e n\cdot A\frac{1}{n\cdot\partial}-e^2n\cdot A\frac{1}{n\cdot\partial}n\cdot A\frac{1}{n\cdot\partial}+i e^3n\cdot A\frac{1}{n\cdot\partial}n\cdot A\frac{1}{n\cdot\partial}n\cdot A\frac{1}{n\cdot\partial}\right. \right.\nn\\
& &\left. \left. +e^4n\cdot A\frac{1}{n\cdot\partial}n\cdot A\frac{1}{n\cdot\partial}n\cdot A\frac{1}{n\cdot\partial}n\cdot A\frac{1}{n\cdot\partial}\right)-M\right)\psi-\frac{1}{4}F_{\mu\nu}F^{\mu\nu},
\eea

\begin{figure}[h!]
\centering
\includegraphics[scale=0.6]{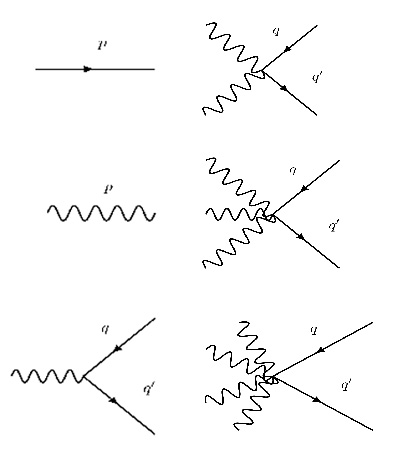}
\caption{VSR Feynman rules with vertex with additional photon lines.}
\label{fig:rules12}
\end{figure}

We use Fourier transform in order to get the Feynman rules in the momenta space. The rules are listed in the table I and the images are in the figure \ref{fig:rules12}.

\begin{table}[h!]
\label{feynman}
\begin{tabular}{|l|l|}
\hline
Electron propagator & $S=i\frac{\left( \slashed{p} + M - \frac{m^2}{2} \frac{\slashed{n}}{n\cdot p}
\right)}{p^2 - M^2_{e}+i\epsilon}$ \\ \hline
Photon propagator  & $P=-i\frac{g_{\mu\nu}}{p^2+i\epsilon}$ \\ \hline
One photon leg Vertex & $V_{1\mu}(q,q')=-i e\left(\gamma_\mu+\frac{1}{2}m^2\frac{\slashed{n}n_\mu}{n\cdot q n\cdot q'}\right)$ \\ \hline
Two photon leg Vertex & $V_{2\mu\nu}(q,q')=-i e^2\frac{1}{2}\slashed{n}m^2\frac{n_\mu n_\nu}{n\cdot q n\cdot q'}\left(\frac{1}{n\cdot(p+q)}+\frac{1}{n\cdot(p'+q)}\right)$ \\ \hline
Three photon leg Vertex & $V_{3\mu\nu\rho}(q,q')=-i e^3\frac{1}{2}\slashed{n}m^2\frac{n_\mu n_\nu n_\rho}{n\cdot q n\cdot q'}\left[\left(\frac{1}{n\cdot(p_3+q)}+\frac{1}{n\cdot(p_2+p_3+q)}\right)+perm. \right]$ \\ \hline
Four photon leg Vertex & $V_{4\mu\nu\rho\sigma}(q,q')=-i e^4\frac{1}{2}\slashed{n}m^2\frac{n_\mu n_\nu n_\rho n_\sigma}{n\cdot q n\cdot q'}\left[\left(\frac{1}{n\cdot(p_4+q)}+\frac{1}{n\cdot(p_3+p_4+q)}+\frac{1}{n\cdot(p_2+p_3+p_4+q)}\right)+perm. \right]$ \\ \hline
\end{tabular}
\caption{Table with the Feynman rules for the Lagrangian in the equation (\ref{lagrangianqedvsr}).}
\end{table}

\section{Photon-Photon Scattering in VSR}
\begin{figure}[h!]
\centering
\includegraphics[scale=0.5]{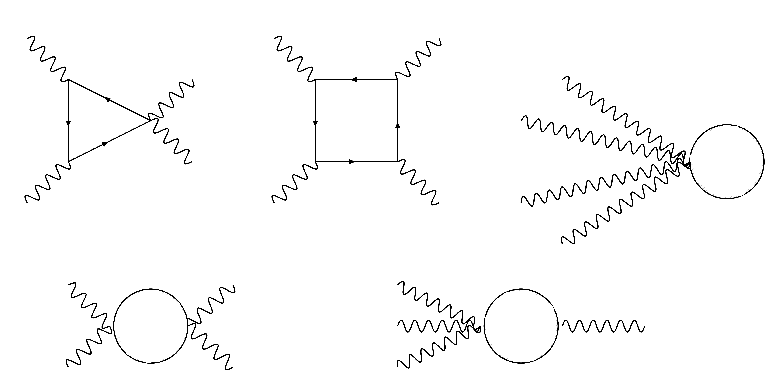}
\caption{The figure shows all the kind of diagrams of four photon legs with fermionic loop in VSR.}
\label{fig:diagrams}
\end{figure}

To compute the photon-photon scattering we consider all the diagrams with four external photon legs, they are five different kinds with its respective permutations. The five diagrams are displayed in the figure \ref{fig:diagrams}.\\ 

We choose one representative by each kind. At the end of the computation we will permute the momenta and sum all of the inequivalent ones to get all the contributions. External photons carry momenta $k_i$, with $i=1,\ldots, 4$. We will consider all the momenta incoming, satisfying $k_1+k_2+k_3+k_4=0$. Thus, working with dimensional regularization, in $d$ dimensions,

\be
\Pi^{\mu \nu \rho \sigma}=\Pi_{sq}^{\mu \nu \rho \sigma}+\Pi_{tri}^{\mu \nu \rho \sigma}+\Pi_{cir}^{\mu \nu \rho \sigma}+\Pi_{three}^{\mu \nu \rho \sigma}+\Pi_{four}^{\mu \nu \rho \sigma}+perm.,
\ee
where

\bea
\Pi_{sq}^{\mu \nu \rho \sigma} &=& \int \frac{d^d p}{(2 \pi)^d}
tr \left[ V^{\mu}_1 (p - k_1, p) S (p - k_1) V^{\nu}_1 (p - k_1 - k_2,
p - k_1) S (p - k_1 - k_2)\times\right. \nn\\
& &\left. V^{\rho}_1 \left( p - k_1 - k_2 - k_3, p - k_1 -
k_2 \right) S (p - k_1 - k_2 - k_3) V^{\sigma}_1 (p, p - k_1 - k_2 - k_3) S
(p) \right],\\
\Pi_{tri}^{\mu \nu \rho \sigma} &=& \int \frac{d^d p}{(2 \pi)^d}
tr [V^{\mu}_1 (p - k_1, p) S (p - k_1) V^{\nu}_1 (p - k_1 - k_2, p -
k_1) S (p - k_1 - k_2) V^{\rho \sigma}_2 (p, p - k_1 - k_2) S (p)],\\
\Pi_{cir}^{\mu \nu \rho \sigma} &=& \int \frac{d^d p}{(2 \pi)^d}
tr [V^{\mu \nu}_2 (p - k_1 - k_2, p) S (p - k_1 - k_2) V^{\rho
\sigma}_2 (p, p - k_1 - k_2) S (p)],\\
\Pi_{three}^{\mu \nu \rho \sigma} &=& \int \frac{d^d p}{(2 \pi)^d}
tr[V^{\mu \nu \rho}_3 (p - k_1 - k_2 - k_3, p) S (p - k_1 - k_2 - k_3)
V^{\sigma}_1 (p, p - k_1 - k_2 - k_3) S (p)],\\
\Pi_{four}^{\mu \nu \rho \sigma} &=& \int \frac{d^d p}{(2 \pi)^d}
tr [V^{\mu \nu \rho \sigma}_4 (p, p) S (p)].
\eea

We will consider the low energy regime. Thus, we expand in Taylor series the electron propagators, keeping only until four $k$ in the numerator, which is equivalent to use the small $k$ approximation.  In addition, we only will keep terms up to $m^2$. The next powers are smaller since $m$ is the neutrino mass, so we can neglect them. From here, the computation is too long to do by hand. We use the software Mathematica and the traces and standard integrations were performed with help of Package-X\cite{Patel:2015tea}.\\

To compute the integrals with $(n\cdot (p+k_i))^{-1}$, where $p$ is the loop momentum to be integrated, we use the Mandelstam-Leibbrandt prescription
\be
\label{M-L}
\frac{1}{n\cdot p}=\frac{\bar{n}\cdot p}{(n\cdot p)(\bar{n}\cdot p)+i\epsilon},
\ee 

where a new null vector $\bar{n}$ is introduced as regulator. This vector satisfies the conditions
\bea
n\cdot\bar{n}=1, \label{cond1}\\
\bar{n}\cdot\bar{n}=0. \label{cond2}
\eea

The integrals are easier to compute following the idea in the reference \cite{Alfaro:2016pjw}, where using the scale symmetry 

\be
\label{n-scale}
n\to\lambda n,\qquad \bar{n}\to\frac{1}{\lambda}\bar{n},\qquad \lambda\in\mathbb{R},
\ee 
the following result has been gotten

\be
\label{ap1}
\int dp\frac{1}{(p^2+2p\cdot q-m^2)^a}\frac{1}{(n\cdot p)^b}=(-1)^{a+b}i\pi^\omega(-2)^b\frac{\Gamma(a+b)}{\Gamma(a)\Gamma(b)}(\bar{n}\cdot q)^b\int_0^1 dt t^{b-1}\frac{1}{(m^2+q^2-2(n\cdot q)(\bar{n}\cdot q)t)^{a+b-\omega}},
\ee
with $\omega=d/2$.\\

With the result in the equation (\ref{ap1}) we can take a derivative with respect to $q$ and we get
\bea
\int d^d p \frac{p_{\mu}}{(p^2 + 2 p \cdot q - m^2)^{a + 1}} \frac{1}{(n
 \cdot p)^b} &=&
(- 1)^{a + b} i \pi^{\omega} (- 2)^{b - 1} \frac{\Gamma (a + b -
\omega)}{\Gamma (a + 1) \Gamma (b)} (\bar{n} \cdot q)^{b - 1} b
\bar{n}_{\mu} \int^1_0 d t t^{b - 1} \frac{1}{[m^2 + q^2 - 2 (n  \cdot q)
(\bar{n} \cdot q) t]^{a + b - \omega}}\nn\\
&+&(- 1)^{a + b} i \pi^{\omega} (- 2)^b \frac{\Gamma (a + b + 1 -
\omega)}{\Gamma (a + 1) \Gamma (b)} (\bar{n}  \cdot q)^b \int^1_0 d t t^{b -
1} \frac{q_{\mu} - t (n  \cdot q \bar{n}_{\mu} + \bar{n}  \cdot q
n_{\mu})}{[m^2 + q^2 - 2 (n  \cdot q) (\bar{n}  \cdot q) t]^{a + b + 1 -
\omega}}.\nn\\
\eea

If we take new partial derivatives respect to $q$ in (\ref{ap1}) we get integrals with more $p$ in the numerator. We use these integrals and the decomposition formula
\be
\frac{1}{(n\cdot(p+k_i))(n\cdot(p+k_j))}=\frac{1}{n\cdot(k_i-k_j)}\left(\frac{1}{(n\cdot(p+k_j)}-\frac{1}{n\cdot(p+k_i)}\right),
\ee
where we have more than one $n$ in the denominator and we perform changes of variables to get always an $(n\cdot p)^{-1}$.\\

After the computation over loop momentum integrals and all the permutations to get $\Pi^{\mu\nu\rho\sigma}$, we have checked it satisfies the Ward identity $\Pi^{\mu\nu\rho\sigma}k_{i\mu}=0$. We will write only the dominant term in $m^2$, which is $m^2/M_e^4$. We will see later that the following term, which contains $m^2/M_e^6$, is zero. Thus, we get an Euler-Heisenberg modified expression
\be
\label{eheq}
\L_{EH}=\frac{e^4}{M_e^4\pi^2}\left[\frac{1}{1440}(F_{\mu\nu}F^{\mu\nu})^2+\frac{7}{5760}(\tilde{F}_{\mu\nu}F^{\mu\nu})^2+\frac{m^2}{8}\left(\frac{\bar{n}^\mu}{n\cdot\partial}\varphi_\mu\right)^2\varphi_\nu\varphi^\nu\right],
\ee
where $\tilde{F}_{\mu\nu}=\frac{1}{2}\epsilon_{\mu\nu\alpha\beta}F^{\alpha\beta}$ is the electromagnetic dual tensor and $\varphi_\nu=n^\mu F_{\mu\nu}$. The first two terms in (\ref{eheq}) are the known standard Euler-Heisenberg terms, while the latter is a new VSR contribution.\\

\section{$\bar{n}$ Prescription and Cross section of the scattering}

The new null vector $\bar{n}$ used to compute the integrals with non local terms breaks the $SIM(2)$ invariance. In the reference \cite{Alfaro:2017umk} for the computation of the vacuum polarization, which diagrams have two external legs, we define

\be
\label{nbark1}
\bar{n}_{\mu} = - \frac{k_1^2}{2 (n \cdot k_1)^2} n_{\mu} +\frac{k_{1\mu}}{n \cdot k_1},
\ee
where $k_1$ is the only independent external momentum, since the relation $k_1+k_2=0$ is satisfied. For the photon-photon scattering, the diagrams have four legs, So, there are three independent momentum, since the relation $k_1+k_2+k_3+k_4=0$ is satisfied.\\

We can define $\bar{n}$ with an arbitrary momentum $p$ as
\be
\bar{n}_{\mu} = - \frac{p^2}{2 (n \cdot p)^2} n_{\mu} +\frac{p_{\mu}}{n \cdot p},
\ee
and it will satisfy the relations (\ref{cond1}) and (\ref{cond2}). If we try to solve the ambiguity using a combination of the momenta, like $k_1+k_2+k_3$ the cross section will present divergences when any momentum $k_i$ coincide with the direction of $n$.\\

To fix it, we will define the Mandelstam-Leibbrandt prescription in the operator sense. Thus,
\be
\label{M-Lop}
\frac{1}{n \cdot \partial} = \frac{\bar{n} \cdot \partial}{n \cdot\partial \bar{n} \cdot \partial + i \varepsilon}.
\ee

In this way, since we are working with operators, we will define $\bar{n}$ as an operator as

\be
\label{nbarop}
\bar{n}_{\mu} = - \frac{\partial^2}{2 (n \cdot \partial)^2} n_{\mu} +\frac{\partial_{\mu}}{n \cdot \partial}.
\ee

This prescription ensures there is only one $\bar{n}$ for all the integrals as requires the Ward Identiy computation. Nevertheless, it is unique in the operator sense, but in the momentum representation it depends on the momentum where it acts. Moreover, we notice this definition satisfies the conditions for $\bar{n}$ stated in the equations (\ref{cond1}) and (\ref{cond2}):
\bea
n \cdot \bar{n} &=& \frac{n^{\mu} \partial_{\mu}}{n \cdot \partial} = 1,\\
\bar{n} \cdot \bar{n} &=& \left( - \frac{\partial^2}{2 (n \cdot\partial)^2} n_{\mu} + \frac{\partial_{\mu}}{n \cdot\partial} \right)
\left( - \frac{\partial^2}{2 (n \cdot \partial)^2} n_{\mu} +
\frac{\partial_{\mu}}{n \cdot \partial} \right) = - \frac{\partial^2}{2 (n
\cdot \partial)^2} - \frac{\partial^2}{2 (n \cdot \partial)^2} +
\frac{\partial^2}{(n \cdot\partial)^2} = 0,
\eea
since $\frac{1}{n \cdot \partial} n \cdot \partial = 1$. Now, we apply it in (\ref{M-Lop}) and we get

\be
\frac{1}{n \cdot \partial}  = \frac{\partial^2}{n \cdot\partial (\partial^2 + i \varepsilon')},
\ee
where $\varepsilon'=2\varepsilon$.\\

We notice this prescription is equivalent to the definition in \cite{Alfaro:2017umk} for the two external legs case, since in that computation there is only one independent momentum available, so the operator $\bar{n}$ applied to the external fields, which have this momentum, is written in the form (\ref{nbark1}) in the momentum space. The operator prescription recovers the two leg case and it is useful for more external legs.\\

Using this prescription in the Euler-Heisenberg equation (\ref{eheq}) we recover the $SIM(2)$ invariance. Hence,
\be
\label{eheq2}
\L_{EH}=\frac{e^4}{M_e^4\pi^2}\left[\frac{1}{1440}(F_{\mu\nu}F^{\mu\nu})^2+\frac{7}{5760}(\tilde{F}_{\mu\nu}F^{\mu\nu})^2+\frac{m^2}{8}\left(\frac{\partial^\mu}{(n\cdot\partial)^2}\varphi_\mu\right)^2\varphi_\nu\varphi^\nu\right],
\ee
where we have used $n^\mu\varphi_\mu=0$.\\

We compute explicitly the unpolarized differential cross section considering all the $m^2$ terms using
\be
\frac{d\sigma}{d\Omega}=\frac{\vert\M\vert^2}{64\pi^2(E_1+E_2)^2},
\ee
where $\vert\M\vert^2$ is defined with
\be
i\mathcal{M}=\Pi^{\mu\nu\rho\sigma}\epsilon_{\mu}(k1)\epsilon_{\nu}(k2)\epsilon^{*}_{\rho}(k3)\epsilon^{*}_{\sigma}(k4),
\ee
and its conjugate. Since our result satisfies the Ward identity, and in consequence it is gauge invariant, we choose the light cone gauge $n\cdot A=0$ to simplify our computations. We can split $\Pi^{\mu\nu\rho\sigma}$ in a standard and a VSR part

\be
\Pi^{\mu\nu\rho\sigma}=\Pi^{\mu\nu\rho\sigma}_{ST}+\Pi^{\mu\nu\rho\sigma}_{VSR}.
\ee

The VSR part can be written as

\be
\Pi^{\mu\nu\rho\sigma}_{VSR}=\Pi^{\mu\nu\rho\sigma}_{VSR4}+\Pi^{\mu\nu\rho\sigma}_{VSR6},
\ee
where 

\bea
\Pi^{\mu\nu\rho\sigma}_{VSR4}&=&\frac{e^4 m^2}{2 M^4_e \pi^2} \frac{1}{n \cdot k_1 n \cdot k_2 n \cdot
k_3 n \cdot (k_1 + k_2 + k_3)} \left( (n \cdot k_1)^4 k_{2 \nu} k_{3 \rho}
g_{\sigma \mu} + 2 (n \cdot k_1)^3 (n \cdot k_2 + n \cdot k_3) k_{2 \nu}
k_{3 \rho} g_{\sigma \mu}\right. \nn\\
& & \left.+ 2 n \cdot k_1 (n \cdot k_2 + n \cdot k_3)
k_{1 \mu} \left( (n \cdot k_2)^2 k_{3 \rho} g_{\sigma \nu} + \left( n
\cdot k_3 \right)^2 k_{2 \nu} g_{\sigma \rho} \right) \right.\nn\\
& &\left. + (n \cdot k_1)^2
\left( 2 n \cdot k_2 n \cdot k_3 k_{2 \nu} k_{3 \rho} g_{\sigma \mu} +
\left( n \cdot k_2 \right)^2 k_{3 \rho} (k_{2 \nu} g_{\sigma \mu} + k_{1
\mu} g_{\sigma \nu} - (k_{1 \sigma} + k_{2 \sigma} + k_{3 \sigma}) g_{\mu
\nu}) \right.\right.\nn\\
& &\left.\left. + (n \cdot k_3)^2 k_{2 \nu} \left( k_{3 \rho} g_{\sigma \mu} + k_{1
\mu} g_{\sigma \rho} - (k_{1 \sigma} + k_{2 \sigma} + k_{3 \sigma}) g_{\mu
\rho} \right) \right) \right.\nn\\
& & \left. + k_{1 \mu} \left( (n \cdot k_2)^4 k_{3 \rho}
g_{\sigma \mu} + 2 (n \cdot k_2)^3 n \cdot k_3 k_{3 \rho} g_{\sigma \nu} +
2 n \cdot k_2 (n \cdot k_3)^3 k_{2 \nu} g_{\sigma \rho} + (n \cdot
k_3)^4 k_{2 \nu} g_{\sigma \rho} \right.\right.\nn\\
& & \left.\left. +(n \cdot k_2)^2 \left( n \cdot k_3
\right)^2 (k_{3 \rho} g_{\sigma \nu} + k_{2 \nu} g_{\sigma \rho} - (k_{1
\sigma} + k_{2 \sigma} + k_{3 \sigma}) g_{\nu \rho}) \right) \right),\\
\Pi^{\mu\nu\rho\sigma}_{VSR6}&=&\frac{1}{6 M^2_e} (k^2_1 + k^2_2 + k^2_3 + k^2_4) \Pi_{VSR4}^{\mu \nu \rho\sigma}.
\eea

We can see $\Pi^{\mu\nu\rho\sigma}_{VSR6}=0$, because the photon satisfies $k_i^2=0$. Thus, the only term is the dominant $\Pi^{\mu\nu\rho\sigma}_{VSR4}$. We use the initial energies $E_1=E_2=\omega$ as usual and we get 

\be
\frac{d\sigma}{d\Omega}=\frac{139\alpha^4\omega^6}{(180\pi)^2M_e^8}(3+\cos^2{\theta})^2,
\ee
where $\alpha=\frac{e^2}{4\pi}\approx\frac{1}{137}$.\\

We notice it is the standard result for the differential cross section in the photon-photon scattering with a mass shift. The extra VSR term does not contributes. This result should not surprise us, since the photon-photon scattering is related with a constant field in the Euler-Heisenberg Lagrangian. Our new term contains a derivative of the field, so, the new term should not appear as the explicit computation shows.




\section{Schwinger proper time method}

We will consider the problem in a different perspective. Here we will use the Schwinger method. We start from the functional
\be
Z [J] = \int \mathcal{D}A_{\mu} \mathcal{D} \bar{\psi} \mathcal{D} \psi \exp
\left\{ i \int d^4 x \bar{\psi} \left( i \slashed{D} - M + i \frac{m^2}{2}
\frac{\slashed{n}}{n \cdot D} \right) \psi - \frac{1}{4} F_{\mu \nu} F^{\mu \nu}\right\}.
\ee

Since the fermionic part is gaussian it is straightforward

\be
\int \mathcal{D} \bar{\psi} \mathcal{D} \psi \exp \left\{ i \int d^4 x
\bar{\psi} \left( i \slashed{D} - M + i \frac{m^2}{2} \frac{\slashed{n}}{n \cdot D}
\right) \psi \right\} = \det \left( i \slashed{D} - M + i \frac{m^2}{2}
\frac{\slashed{n}}{n \cdot D} \right).
\ee

We can express the determinant as





\be
\ln \left(\det \left( i \slashed{D} - M + i \frac{m^2}{2} \frac{\slashed{n}}{n
\cdot D} \right) \right) = \frac{1}{2} Tr \left\{ \ln \left( - \left(
i \slashed{D} + i \frac{m^2}{2} \frac{\slashed{n}}{n \cdot D} \right)^2 + M^2
\right) \right\}.
\ee







Here, $Tr$ is the trace in the space as well as of the gamma matrices. We rearrange the first term in the logarithm and after some manipulations we get

\be
\ln \left( \det \left( i \slashed{D} - M + i \frac{m^2}{2} \frac{\slashed{n}}{n
\cdot D} \right) \right) = \frac{1}{2} Tr \left\{ \ln \left( D^2 +
M_e^2 + \sigma^{\mu \nu} \left( \frac{e}{2} F_{\mu \nu} - i \frac{m^2}{2}
\left[ D_{\mu}, \frac{n_{\nu}}{n \cdot D} \right] \right) \right) \right\},
\ee

where $\sigma^{\mu \nu}=\frac{i}{2} [\gamma^{\mu}, \gamma^{\nu}]$. We notice it appears a commutator $\left[ D_{\mu}, \frac{n_{\nu}}{n \cdot D} \right]$.\\

We will use the Schwinger prescription
\be
\ln A = - \int^{\infty}_0 \frac{d s}{s} e^{i s A}.
\ee
Thus, after the change $s\to -s$ and writing explicitly the trace in $x$, keeping only $tr$, which is the trace in the gamma matrices:
\be
\ln \left[ \det \left( i \slashed{D} - M + i \frac{m^2}{2} \frac{\slashed{n}}{n
\cdot D} \right) \right] = \frac{1}{2} \int^{\infty}_0 \frac{d s}{s} e^{- i
s M^2_e} \int d^4 x tr [\langle x | e^{- i s H} | x \rangle],
\ee

with 
\be
\label{hamspt}
H = D^2 + \sigma^{\mu \nu} \left( \frac{e}{2} F_{\mu \nu} - i \frac{m^2}{2}\left[ D_{\mu}, \frac{n_{\nu}}{n \cdot D} \right] \right).
\ee

In this case, $H$ acts as a hamiltonian with proper time $s$. Here, the new VSR term acts like a potential and we notice when $m\to0$ we recover the standard case.\\

We compute the commutator writing the non-local term as an integral

\be
\left[ D_{\mu}, \frac{1}{n \cdot D} \right] = \int^{\infty}_0 d \eta[D_{\mu}, e^{- \eta n \cdot D}].
\ee
We expand the exponential and using $[D_\mu, D_\nu]=i e F_{\mu\nu}$, we get

\bea
\left[ D_{\mu}, \frac{1}{n \cdot D} \right] &=& i e \int^{\infty}_0 d \eta
\left( - \eta n^{\rho} F_{\mu \rho} + \frac{\eta^2}{2} n^{\rho} n^{\lambda}
F_{\mu \rho} D_{\lambda} + \frac{\eta^2}{2} n^{\rho} n^{\lambda} D_{\rho}
F_{\mu \lambda} - \frac{\eta^3}{3!} n^{\rho} n^{\lambda} n^{\kappa} F_{\mu
\rho} D_{\lambda} D_{\kappa}- \frac{\eta^3}{3!} n^{\rho} n^{\lambda}
n^{\kappa} D_{\rho} D_{\lambda} F_{\mu \kappa}\right.\nn\\
& &\left.  - \frac{\eta^3}{3!} n^{\rho}
n^{\lambda} n^{\kappa} D_{\rho} F_{\mu \lambda} D_{\kappa} + \frac{\eta^4}{4!}
n^{\rho} n^{\lambda} n^{\kappa} n^{\delta} F_{\mu \rho} D_{\lambda} D_{\kappa}
D_{\delta} + \frac{\eta^4}{4!} n^{\rho} n^{\lambda} n^{\kappa} n^{\delta}
D_{\rho} F_{\mu \lambda} D_{\kappa} D_{\delta}\right. \nn\\
& &\left. + \frac{\eta^4}{4!} n^{\rho}
n^{\lambda} n^{\kappa} n^{\delta} D_{\rho} D_{\lambda} F_{\mu \kappa}
D_{\delta} + \frac{\eta^4}{4!} n^{\rho} n^{\lambda} n^{\kappa} n^{\delta}
D_{\rho} D_{\lambda} D_{\kappa} F_{\mu \delta} - \cdots \right).
\eea

We will move all the $F$ to the right using the commutator $[D_\lambda,F_{\mu,\rho}]$ and nested commutators. Hence,

\bea
\left[ D_{\mu}, \frac{1}{n \cdot D} \right] &=& i e \int^{\infty}_0 d \eta
\eta \left( 1 - \eta n^{\lambda} D_{\lambda} + \frac{\eta^2}{2} n^{\rho}
n^{\lambda} D_{\rho} D_{\lambda} - \frac{\eta^3}{3!} n^{\rho} n^{\lambda}
n^{\kappa} D_{\rho} D_{\lambda} D_{\kappa} + \cdots \right) \varphi_{\mu}\nn\\
& & + i e \int^{\infty}_0 d \eta \left( - \frac{\eta^2}{2} n^{\rho} n^{\lambda}
[D_{\lambda}, F_{\mu \rho}] + 3 \frac{\eta^3}{3!} n^{\rho} n^{\lambda}
n^{\kappa} D_{\lambda} \left[ D_{\kappa}, F_{\mu \rho} \right] - 6
\frac{\eta^4}{4!} n^{\rho} n^{\lambda} n^{\kappa} n^{\delta} D_{\lambda}
D_{\delta} [D_{\kappa}, F_{\mu \rho}] + \cdots \right)\nn\\
& & + i e \int^{\infty}_0 d \eta \left( - \frac{\eta^3}{3!} n^{\rho} n^{\lambda}
n^{\kappa} [D_{\kappa}, [D_{\lambda}, F_{\mu \rho}]] + \frac{\eta^4}{3!}
n^{\rho} n^{\lambda} n^{\kappa} n^{\delta} D_{\lambda} \left[ D_{\delta},
[D_{\kappa}, F_{\mu \rho}] \right]\right. \nn\\
& & \left.- \frac{\eta^4}{4!} n^{\rho} n^{\lambda}
n^{\kappa} n^{\delta} [D_{\delta}, [D_{\kappa}, [D_{\lambda}, F_{\mu \rho}]]]
+ \cdots \right).
\eea

Rearranging the terms we get
\be
\left[ D_{\mu}, \frac{1}{n \cdot D} \right] = i e \int^{\infty}_0 d \eta
e^{- \eta n \cdot D} \left( \eta \varphi_{\mu} + \frac{\eta^2}{2} [n \cdot
D, \varphi_{\mu}] + \frac{\eta^3}{3!} [n \cdot D, [n \cdot D,
\varphi_{\mu}]] + \frac{\eta^4}{4!} [n \cdot D, [n \cdot D, [n \cdot D,
\varphi_{\mu}]]] + \ldots \right).
\ee

Using the gamma function
\be
\int^{\infty}_0 d \eta \eta^m e^{- \eta n \cdot D} = \frac{m!}{(n \cdot D)^{m + 1}},
\ee
thus,
\be
\left[ D_{\mu}, \frac{1}{n \cdot D} \right] = i e \left( \frac{1}{(n
\cdot D)^2} \varphi_{\mu} + \frac{1}{(n \cdot D)^3} [n \cdot D,
\varphi_{\mu}] + \frac{1}{(n \cdot D)^4} [n \cdot D, [n \cdot D,
\varphi_{\mu}]] + \frac{1}{(n \cdot D)^5} [n \cdot D, [n \cdot D, [n
\cdot D, \varphi_{\mu}]]] + \ldots \right).
\ee

The commutator $\left[ D_{\mu}, \frac{1}{n \cdot D} \right]$ is an infinite serie of nested commutators. In this way an exact computation is too hard to manage. However, if we use a constant field $F$ to get the Euler-Heisenberg terms, it is easy to see that

\be
\label{commut}
\left[ D_{\mu}, \frac{1}{n \cdot D} \right] =0.
\ee

Replacing (\ref{commut}) in (\ref{hamspt}) we get

\be
H = D^2 + \frac{e}{2} \sigma^{\mu \nu} F_{\mu \nu},
\ee

which is the standard result. It implies that the VSR terms does not contributes to the photon-photon scattering cross section, as we stated in the last section using a diagrammatic point of view.

\section{Summary and conclusions}

We have presented an operator prescription for $\bar{n}$ to manage the integrals in VSR computations. This result is useful to treat any perturbative calculation in the model, where the easiest way to proceed has been to use dimensional regularization and the integrals computed in \cite{Alfaro:2016pjw}. This result is valid for any diagram with an arbitrary number of external legs. Our definition in the momentum space is the same Mandelastam-Leibbrandt prescription and it recovers the prescription of \cite{Alfaro:2017umk} in the two external legs case. We have applied the new prescription in the computation of the low energy regime in the photon-photon scattering. We observe with this prescription the result is divergence-free and here we have recovered the standard result for the unpolarized differential cross section.\\

Since the low energy limit in the photon-photon scattering is related with a constant field, the diagrammatic and the Schwinger method shows the only modification of the VSR terms is a mass shift. In addition, the on-shell condition $k_i^2=0$ implies that the VSR effects vanishes. For the vacuum polarization the result given in \cite{Alfaro:2016pjw} is

\be
\Pi_{\mu \nu} = A (q^2 g_{\mu \nu} - q_{\mu} q_{\nu})+B\left(- q^2 \frac{n_{\mu}
n_{\nu}}{(n \cdot q)^2}+\frac{q_{\mu} n_{\nu} + q_{\nu} n_{\mu}}{n \cdot
q} - g_{\mu \nu}\right),
\ee
where

\bea
A &=& -e^2 \frac{i}{(4 \pi)^{\omega}} \int^1_0 d x \Gamma (2 - \omega) \frac{8 x
(1 - x)}{(M^2_e - (1 - x) x q^2)^{2 - \omega}},\\
B&=&m^2 i \frac{e^2}{4 \pi^2} \int^1_0 d x
\frac{1}{(1 - x)} \log \left( 1 + \frac{(1 -
x)^2 q^2}{M^2_e - (1 - x) q^2} \right).
\eea
Notice where we set $q^2=0$, $B$ vanishes and we recover the same standard result. The same occurs in the photon-photon scattering. This results seems to indicate that the mass shift is the only effect of the VSR terms in the gauge bosons processes. VSR modifications and possible signals of the privileged directions should appear when a fermion is involved in the processes, as in the Coulomb scattering\cite{Alfaro:2019koq} or Bhabha and Compton scattering\cite{Bufalo:2019kea}.

\begin{acknowledgements}

The work of A.Soto is supported by the CONICYT-PFCHA/Doctorado Nacional/2017-21171194 and
Fondecyt 1150390. The work of J. Alfaro is partially supported by Fondecyt 1150390 and CONICYT-PIA-ACT1417.\\

\end{acknowledgements}
\vspace{-20pt}




\end{document}